\documentclass[a4paper,conference]{IEEEtran}

\usepackage{graphicx}
\usepackage{url}
\usepackage{placeins}
\usepackage{booktabs}
\usepackage{subfigure}
\usepackage{color}
\usepackage{amsmath}
\usepackage[first,light,timestamp]{draftcopy}

\begin{document}

\title{Secure Vehicular Communication Systems: \\ Design and Architecture}

\author{P. Papadimitratos, L. Buttyan, T. Holczer, E. Schoch, J. Freudiger, M. Raya \\
Z. Ma, F. Kargl, A. Kung, J.-P. Hubaux}

\maketitle

\begin{abstract}
Significant developments took place over the past few years in the area of vehicular communication (VC) systems. Now, it is well-understood in the community that security and protection of private user information are a prerequisite for the deployment of the technology. This is so exactly because the benefits of VC systems, with the mission to enhance transportation safety and efficiency, are at stake. Without the integration of strong and practical security and privacy enhancing mechanisms, VC systems could be disrupted or disabled even by relatively unsophisticated attackers. We address this problem within the \emph{SeVeCom} project, having developed a security architecture that provides a comprehensive and practical solution. We present our results in a set of two papers in this issue. In this first one, we analyze threats and types of adversaries, we identify security and privacy requirements, and present a spectrum of mechanisms to secure VC systems. We provide a solution that can be quickly adopted and deployed. Our progress towards implementation of our architecture, along with results on the performance of the secure VC system, are presented in the second paper. We conclude with an investigation, based on current results, of upcoming elements to be integrated in our secure VC architecture.
\end{abstract}

\section{Introduction}

After the deployment of various vehicular technologies, such as toll collection or active road-signs, vehicular communication (VC) systems are emerging. They comprise network nodes, that is, vehicles and road-side infrastructure units (RSUs), equipped with on-board
sensory, processing, and wireless communication modules. Vehicle-to-vehicle (V2V) and vehicle-to-infrastructure (V2I) communication can enable a range of applications to enhance transportation safety and efficiency, as well as infotainment. For example, they can send warnings on environmental hazards (e.g., ice on the pavement), traffic and road conditions (e.g., emergency braking, congestion, or construction sites), and local (e.g., tourist) information.

The unique features of VC are a double-edged sword: a rich set of tools will be available but a formidable set of abuses and attacks becomes possible. Consider, for example, an attacker that ``contaminates'' large portions of the vehicular network with false information: a single compromised vehicle can transmit false hazard warnings, which can then be taken up by all vehicles in both traffic streams. Or a tampered vehicle that forges messages to masquerade as an emergency vehicle to mislead other vehicles to slow down and yield. Or a different type of attacker, which deploys a number of receivers, records messages transmitted by the vehicles, especially safety beacons which report the vehicle's location, to track a vehicle's location and transactions, and infer private information about its driver and passengers.

It is clear that to thwart such attacks, security and privacy-enhancing mechanisms are necessary; in fact, they are a prerequisite for deployment. Otherwise VC systems could make anti-social and criminal behavior easier, in ways that would actually jeopardize the benefits of their deployment. This has been recently well understood in academia, the industry, and among authorities; and a number of concerted efforts have been undertaken to design security architectures for VC systems.

A prominent example of those efforts is our three-year European-funded \emph{Secure Vehicular Communications} (SeVeCom) Project (http://www.sevecom.org), which is approaching its conclusion at the end of 2008. In this project, universities, car manufacturers, and car equipment suppliers collaborate on the design of a baseline architecture that provides a level of protection sought by users and legislators and is practical. Our baseline architecture is based on well-established and understood cryptographic primitives but can also be tuned or augmented, to meet more stringent future requirements.

In this paper, we first discuss the capabilities of attackers. Then, we present the requirements based on which we develop our architecture. The basic aspects we seek to address are: identity and cryptographic key management, privacy protection, secure communication, and in-car protection. Next, we provide details on credential management and cryptographic support, which enable secure and privacy-enhancing communication. We conclude with a short discussion that ushers our second article, which is concerned with implementation and performance issues, and upcoming research challenges.

\section{Adversary Model}\label{sec:adversary}

VC system entities can be \emph{correct} or \emph{benign}, that is, comply with the implemented protocols, or they may deviate from the protocol definition, that is, be \emph{faulty} or \emph{adversarial}.  Adversarial behavior can vary widely, according to the implemented protocols and the capabilities of the adversary. Its incentive may be own benefit or malice. We do not consider here benign faults, for example, communication errors, message delaying or loss, which can occur either under normal operational conditions or due to equipment failure. Instead, we focus on adversarial behavior, which can cause a much larger set of faults. We do not dwell on individual VC protocols for which to describe attacks. Rather, we survey the capabilities of adversaries and discuss aspects relevant to the VC context. A more detailed exposition, which also discusses models used in other types of distributed systems, is available in \cite{PapadGH:06}.

Even though the VC protocol implementations will be proprietary, open definitions of standards will provide attackers with detailed knowledge about the system operation. Any wireless device that runs a rogue version of the VC protocol stack poses a threat. Attackers can either be passive or active.

\emph{Active adversaries} can meaningfully \emph{modify} in-transit messages they relay, beyond the modifications the protocol definitions allow or require them to perform. Or, more generally, they can \emph{forge}, that is, synthesize in a  manner non-compliant to the protocols and system operation, and \emph{inject} messages. Since adversaries are aware of the VC protocols, they can choose any combination of these actions according to their own prior observations (messages they received) and the protocol they attempt to compromise. An active adversary may also \emph{jam} communications, that is, interfere deliberately and prevent other devices within its range to communicate. Or, it can \emph{replay} messages that it received and were previously transmitted by other system entities. In contrast to active adversaries, \emph{passive attackers} only learn information about system entities and cannot affect or change their behavior.

It is important to distinguish adversaries equipped with cryptographic keys and credentials that entitle them to participate in the execution of the VC system protocols. We denote those as \emph{internal} adversaries. In contrast, adversaries that do not possess such keys and credentials are \emph{external}. We emphasize that the possession of credentials does not guarantee correct operation of the nodes. For example, the \emph{on-board units} (OBUs) can be tampered with and their functionality modified (e.g., by installing a rogue version of the protocol stack). Or, the cryptographic keys of an RSU or a vehicle can be compromised (e.g., physically extracted from an unattended vehicle) and be utilized by an adversarial device. If this were the case, a node with multiple (compromised) keys could appear as multiple nodes.

More generally, \emph{multiple adversarial nodes} can be present in the network at different locations. They can be acting \emph{independently} or they may \emph{collude}, i.e., exchange information and coordinate their actions, in order to mount a more effective attack.\footnote{We emphasize though that even in that case, adversaries are \emph{computationally limited} and unable to break keys of other nodes.} For example, they could all report an imaginary event (e.g., traffic jam or accident), in order to mislead correct nodes to think this is indeed the case. Over time, the set of adversarial nodes can change both in numbers and locations. On the one hand, the compromised nodes, for example, illegally modified vehicles, can increase over time, as drivers may have some benefit in doing so. On the other hand, fault detection mechanisms and diagnostics, along with policy enforcement can lead to gradual eradication of faulty devices.

Overall, however, it is reasonable to expect that only a \emph{relatively small fraction} of the VC devices will be adversaries. Of course this depends on the appropriate design of the system, which should not allow for easy exploits (e.g., malware propagation). Moreover, the majority of the users will not have the expertise and the motivation to tamper with their VC devices, while maintenance will address the majority of equipment faults.

Given a small fraction of faulty (adversarial) devices, the adversary will have overall \emph{limited physical presence}. Since the transmission range of faulty devices cannot be unbounded, even if they had customized hardware that exceeds the communication range of vehicular or road-side devices, the adversaries can affect only a fraction of the VC system area. Within this area, they can cause denial of service and do it in a selective manner, i.e., erase one or more messages sent by other nodes. This does not preclude that a few adversarial devices surround a correct node (vehicle) at some point in time. But most often and in most locations, correct nodes will encounter few or only a single adversary.

Due to the nature of VC systems, with vehicles equipped with a number of sensors, exchange of false measurements can compromise the VC-enabled applications. An arguably convenient attack, in the sense that it may be relatively easy to mount, is by \emph{controlling the sensory inputs} to the OBU instead of attempting to compromise the OBU or its cryptographic keys. Tampering with a sensor or with the OBU-sensor connection may indeed be simpler. It is not easy to classify an \emph{input-controlling adversary} as external or internal. On the one hand, no access to credentials and cryptographic material is necessary. On the other hand, messages generated and transmitted due to the input-controlling adversary originate from a legitimate system participant. What we should note though is that such an adversary is relatively weaker than an internal one: controlling inputs alone cannot induce arbitrary behavior, if self-diagnostics and other controls are available and out of reach of the adversary. 

\section{Security Requirements}\label{sec:sec-req}

The problem at hand is to secure the operation of VC systems, that is, design protocols that mitigate attacks and thwart deviations from the implemented protocols to the greatest possible extent. Different protocols have their own specifications, that is, sought properties. Rather than providing an exhaustive enumeration of requirements per protocol and application, we identify first a set of stand-alone requirements. Then, we outline a number of example VC applications along with the related security requirements.

The identified stand-alone security requirements are the following:

\textbf{Message Authentication and Integrity}, to protect against any alteration and allow the receiver of a message to corroborate the sender of the message.

\textbf{Message Non-Repudiation}, so that the sender of a message cannot deny having sent a message.

\textbf{Entity Authentication}, so that a receiver is ensured that the sender generated a message \emph{and} has evidence of the \emph{liveness} of the sender. In other words, ascertain that a received unmodified message was generated within an interval $[t-\tau, t]$, with $t$ the current
time at the receiver and $\tau>0$ a sufficiently small positive value.

\textbf{Access Control}, to determine via specific system-wide policies the assignment of distinct roles to different types of nodes and their allowed actions within the system. As part of access control, \textbf{authorization} establishes what each node is allowed to do in the network, e.g., which types of messages it can insert in the network, or more generally the protocols it is allowed to execute.

\textbf{Message Confidentiality}, to keep the content of a message secret from those nodes not authorized to access it.

\textbf{Accountability}, to be able to map security-related events to system entities.

\textbf{Privacy Protection}, to safeguard private information of the VC system users. This is a general requirement that relates to the protection of private information stored off-line. In the context of communication, which is the object of SeVeCom, we are interested in \textbf{anonymity} for the actions (messages and transactions) of the vehicles. We elaborate on the VC-specific aspects that we seek to address next.

For privacy, along with security, we focus on private vehicles (e.g., excluding emergency vehicles, buses, etc). This is so, as the operation of all other VC nodes, including RSUs, does not raise any privacy concerns, and all those other nodes should be readily identifiable. A primary concern for VC systems is to provide \emph{location privacy}, that is, prevent others (any \emph{observer}) from learning past or future locations of a VC system user (vehicle driver or passenger). With our focus on VC, we can safeguard location privacy by seeking to satisfy a more general requirement, anonymity for the vehicle message transmissions.

Ideally, it should be impossible for any observer to learn if a specific vehicle transmitted or will transmit in the future a message (more generally, take an action, that is, be involved in a VC protocol), and it should be impossible to link any two or more messages (in general, actions) of the same vehicle. Even if an observer tried to guess, that should leave only a low probability of linking a vehicle's actions or identifying it among the set of all vehicles, the \emph{anonymity set}. We will elaborate on this notion when we discuss below the management of identities and credentials for VC system entities.


Rather than aiming for this strong anonymity, we require a relatively weaker level of protection: messages should not allow the identification of their sender, and two or more messages generated by the same vehicle should be difficult to link to each other. More precisely, messages produced by a vehicle over a protocol-selectable period of time $\tau$ can always be linked by an observer that received them. But messages $m_1, m_2$ generated at times $t_1, t_2$ such that $t_2 > t_1 + \tau$ cannot. In terms of the observer, we assume that its physical presence is bounded, as stated earlier for the adversary.

In addition, features that enhance \textbf{availability} are required, to enable protocols and services to remain operational even in the presence of faults, malicious or benign. This implies resilience to resource depletion attacks, as well as self-stable protocols which resume their normal operation after the ``removal'' of the faulty participants.

Based on these considerations, SeVeCom performed a detailed requirements analysis where general application characteristics and security requirements were assessed for a large number of VC applications~\cite{KarglEtAl:2006:SecurityEngineeringVANETs}. Table~\ref{tab:requirements} shows a small excerpt from this analysis, with higher values indicating higher importance of a given requirement. For example, for a work zone warning message, it may be relatively less important to rigidly determine its recency. For a collision avoidance application though, it is crucial to ensure the message recency. Of course, for both applications, it is critical to ensure that no message content is fabricated by an attacker. Regarding privacy protection, this is not required for infrastructure- or public vehicle-sent messages, as are the work zone and emergency vehicle warnings.

\begin{table}
\center
\begin{tabular}{|l|c|c|c|c|c|c|}
\hline
 & \multicolumn{3}{|c|}{Feature} & \multicolumn{3}{|c|}{Requirement} \\
\hline
\textbf{Application} &
\rotatebox{90}{\textbf{Safety Appl.}} &
\rotatebox{90}{\textbf{V2V / V2I}} &
\rotatebox{90}{\textbf{Multi-Hop}} &
\rotatebox{90}{\textbf{Authentication  }} &
\rotatebox{90}{\textbf{Integrity}} &
\rotatebox{90}{\textbf{Privacy}} \\
\hline
Intersection collision warning & $\surd$ & V2V & & 2 & 2 & 2 \\
Emergency vehicle signal & $\surd$ & V2I & $\surd$ & 2 & 2 & 0\\
Work zone warning & $\surd$ & V2I & $\surd$ & 1 & 2 & 0 \\
Forward collision warning & $\surd$ & V2V & $\surd$ & 2 & 2 & 2 \\
Cooperative adaptive cruise control & & V2V & $\surd$ & 2 & 2 & 2 \\
\hline
\end{tabular}
\caption{\label{tab:requirements} Sampled VC applications: features and importance of security requirements.}
\end{table}

\section{Secure VC System Overview}\label{sec:svc}

\begin{figure}[t]
\begin{center}
\includegraphics[width = \columnwidth]{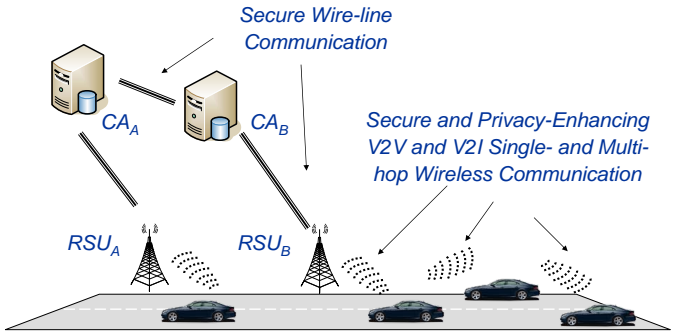}
\caption{Abstract View of the Secure Vehicular Communication System.} \label{fig:abstract-view}
\end{center}
\end{figure}

Our architecture addresses the following fundamental issues: (i) \emph{identity, credential, and key management}, and (ii) \emph{secure communication}. We focus primarily on securing the operation of the wireless part of the VC system, and enhancing the privacy of its users, seeking to satisfy the requirements we outlined earlier in this article. We are fully aware of the projected co-existence of VC-specific and TCP/IP protocol stacks in VC systems. Moreover, towards further strengthening our architecture, we have investigated and developed approaches to address \emph{in-car protection} and \emph{data consistency}, discussed in \cite{KarglPH-:08}. An abstract view of the secure VC system, with \emph{nodes} (vehicles and RSUs) and \emph{authorities} ($CA_A$ and $CA_B$), is shown in Fig.~\ref{fig:abstract-view}. We outline next the main elements of our architecture.

\textbf{Authorities} Drawing from the analogy with existing administrative processes and automotive authorities (e.g., city or state transit authorities), we assume that a large number of \emph{Certification Authorities} (CAs) will be instantiated. Each CA is responsible for a \emph{region} (national territory, district, county, etc.) and manages identities and credentials of all nodes registered with it. To enable interactions between nodes from different regions, CAs provide certificates for other CAs (cross-certification) or provide \emph{foreigner certificates} to vehicles that are register with another CA when they cross the geographical boundaries of their region \cite{PapMH2008}.

\begin{figure}[t]
\begin{center}
\includegraphics[width = \columnwidth]{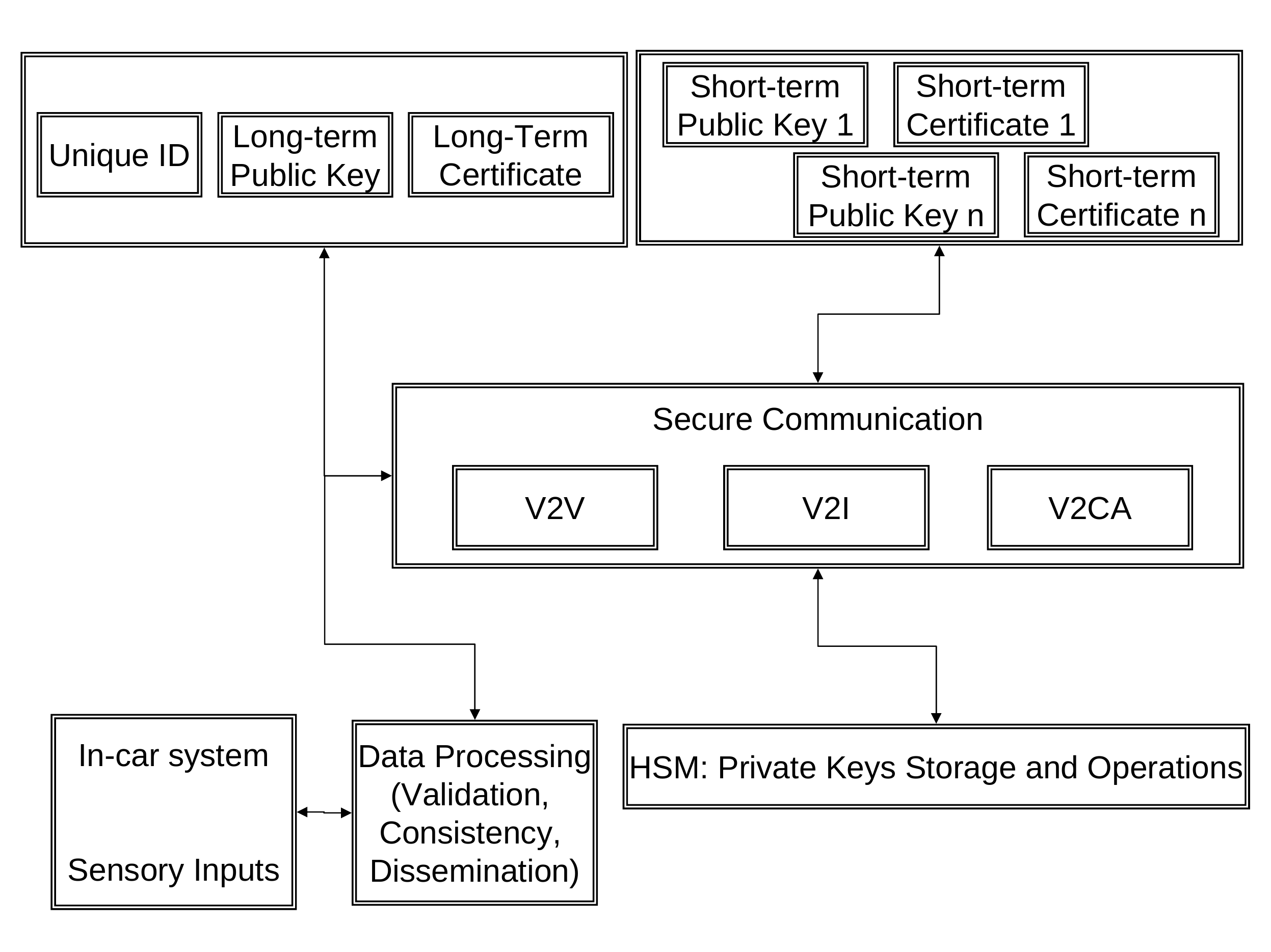}
\caption{Conceptual Secure VC Architecture View: Node functionality.} \label{fig:node-view}
\end{center}
\end{figure}

\textbf{Node Identification} Each node is registered with only one CA, and has a unique \emph{long-term} identity and a pair of \emph{private} and \emph{public} cryptographic keys, and it is equipped with a long-term \emph{certificate}. A list of \emph{node
attributes} and a \emph{lifetime} are included in the certificate, which the CA issues upon node registration and upon certificate expiration. The CA is also responsible for the \emph{eviction} of nodes or the \emph{withdrawal} of compromised cryptographic keys via revocation of the corresponding certificates. In all cases, the interaction of nodes with the CA is infrequent and intermittent, with the road-side infrastructure acting as a gateway to and from the vehicular part of the network, with the use of other infrastructure (e.g., cellular) also possible. The conceptual view of VC nodes is illustrated in Fig.~\ref{fig:node-view}. The node identity and credential management and the role of the HSM, methods to secure V2V and V2I communication, and CA-vehicle interactions (V2CA) that include the issuance of \emph{short-term credentials} to secure vehicle transmissions, are discussed in the rest of the paper. The in-car system and data processing functionality are discussed in \cite{KarglPH-:08}.

\textbf{Hardware Security Module (HSM)} We envision that both vehicles and RSUs are equipped with an HSM, whose purpose is to store and physically protect sensitive information and provide a secure time base. This information is primarily private keys for signature generation. If modules were to be tampered with, to extract private keys, the physical protection of the unit would ensure that the sensitive information (private keys) would be erased, thus preventing the adversary from obtaining them. In addition, the HSM performs all private key cryptographic operations with the stored keys, in order to ensure that sensitive information never leaves the physically secured HSM environment. Essentially, the HSM is the basis of trust; without it, private keys could be compromised and their holders could masquerade as legitimate system nodes.

\textbf{Secure Communication} Digital signatures are the basic tool to secure communications, used for all messages. To satisfy both the security and anonymity requirements, we rely on a \emph{pseudonymous authentication} approach. Rather than utilizing the same long-term public and private key for securing communications, each vehicle utilizes multiple \emph{short-term} private-public key pairs and certificates. A mapping between the short-term credentials and the long-term identity of each node is maintained by the CA.

The basic idea is that (i) each vehicle is equipped with multiple certified public keys (pseudonyms) that do not reveal the node identity, and (ii) the vehicle uses each of them for a short period of time, and then switches to another, not previously used pseudonym. This way, messages signed under different pseudonyms cannot be linked. Signatures, calculated over the message payload, a time-stamp and the coordinates of the sender, can be generated by the originator of a message as well as relaying nodes, depending on the protocol functionality. We provide security for frequently broadcasted \emph{beacon} messages for safety, \emph{restricted flooding} of messages within a geographical region or a hop-distance from the sender, and \emph{position-based routing} used to transmit messages through a single route of relay nodes, where the nodes select as next hop their neighbor with minimum remaining geographical distance to the destination position.

\section{Credential Management and Cryptographic Support}\label{sec:id-crypto}

The management of credentials, both short and long-term, is undertaken by the CAs, which are also responsible for the revocation of credentials for any node if needed, as well as holding the node accountable, by mapping node communications to its long-term identity. Public key operations are performed by the OBU, but all private key operations are performed by the HSM, which is essentially the trusted computing base of the secure VC system.



\subsection{Identity and Credential Management}

\subsubsection{Long-Term Identification}\label{sec:ltid}

Each node $X$ has a unique long-term identity $\mathit{ID}_X$, which will be the outcome of an agreement between car manufacturers and authorities, similar to the use of Vehicle Identification Numbers (VINs). Identifiers of the same format will be assigned both to vehicles and road-side units. Each identity is associated with a cryptographic key pair $(\mathit{SK}_X, \mathit{PK}_X)$, and a set of attributes of node $X$. The attributes reflect technical characteristics of the node equipment (for example, type, dimensions, sensors and computing platform), as well as the role of the node in the system. Nodes can be, for example, private or public vehicles (buses), or vehicles with special characteristics (police patrol cars), or RSUs, with or without any special characteristics (offering connectivity to the Internet). The assignment of an identity, the selection of attributes appropriate for each node, and the generation of the certificate are performed ``off-line,'' at the time the node is registered with the CA. The lifetime of the certificate is naturally long, following the node life-cycle (or a significant fraction of it).

\subsubsection{Short-Term Identification}\label{sec:stid}

To obtain pseudonyms, a vehicle $V$'s HSM generates a set of key pairs $\{(SK^1_V, PK^1_V), ..., (SK^i_V, PK^i_V)\}$ and sends the public keys to a corresponding CA via a secured communication channel. $V$ utilizes its long-term identity $ID_V$ to authenticate itself to the CA. The CA signs each of the public keys, $PK^i_V$, and generates a set of pseudonyms for $V$. Each pseudonym contains an identifier of the CA, the lifetime of the pseudonym, the public key, and the signature of the CA; thus, no information about the identity of the vehicle.

Pseudonyms are stored and managed in the on-board pseudonym pool, with their corresponding secret keys kept in the HSM. This ensures that each vehicle has exactly one key pair (own pseudonym pseudonym and private key) that is active during each time period. Moreover, once the switch from the $(SK_j,PK_j)$ to the $j+1$-st key pair $(SK_{j+1},PK_{j+1})$ is done, no messages can be further signed with $SK_j$; even if the certificate for $PK_j$ is not yet expired. In other words, pseudonymity cannot be abused: For example, a rogue vehicle cannot sign multiple beacons each with a different $SK_j$ over a short period, and thus cannot appear as multiple vehicles.\footnote{The CA could prevent abuse of the pseudonymity by issuing short-term certificates with non-overlapping lifetimes. We also note that multiple pseudonyms can active simultaneously only when used for completely disjoint communication (e.g., one for all safety messaging and one for infotainment downloads.}

A vehicle needs to contact the CA, infrequently but regularly, to obtain a new set of pseudonyms. For example, if a vehicle utilizes pseudonyms in set $i$, it obtains the $(i+1)$-st set of pseudonyms while it can still operate with the $i$-th set. It switches to the $(i+1)$-st set once no pseudonym in the $i$-th set can be used. We term this process a \textit{pseudonym refill}.

Due to the requirement for accountability, the CA archives the issued pseudonyms together with the vehicle's long-term identity. In case of an investigation, an authorized party can ask the CA to perform a \textit{pseudonym resolution}: Reveal the link of a specific pseudonym to the long-term identity of the vehicle (this leading further to its registered owner).

By using the same pseudonym only for a short period of time and switching to a new one, vehicle activities can be only linked over the period of using the same pseudonym. Changing pseudonyms makes it difficult for an adversary to link messages from the same vehicle and track its movements. However, the inclusion of the identity of the $CA_A$ issuing the credential (pseudonym) implies that the vehicle is part of the set of all vehicles registered with $CA_A$. In fact, this is the anonymity set of vehicle $V$. This implies that, for example, a Swiss vehicle should be anonymous within the set of all Swiss vehicles.

This division of vehicles into disjoint subsets, one per CA, allows an observer to rule out a significant portion of vehicles given geographical constraints. Consider again a Swiss vehicle, driving in the East Balkans where it is not likely to encounter numerous other vehicles with the same registration. An observer could then be successful with high probability in guessing that all Swiss pseudonyms (and thus associated messages) are used by the same Swiss vehicle. To prevent such inferences, we require that vehicles crossing the boundaries of a foreign region, $B$, obtain short-term credentials from the local $CA_B$ \cite{PapMH2008}. In our example, $V$ would have to first prove to $CA_B$ it is registered with $CA_A$, then obtain pseudonyms by $CA_B$, and use them exclusively while in region $B$. This way, it would avoid ``standing out'' in region $B$, appearing to any observer of the VC system traffic as part of the anonymity set $B$.


\subsection{Hardware Security Module}\label{sec:hsm}

The Hardware Security Module (HSM) is the trusted computing base of the SeVeCom security architecture. It stores the private cryptographic key material, and provides cryptographic functions to be used by other modules. The HSM is physically separated from the On-Board Unit (OBU), and it has some tamper resistant properties in order to protect the private key material against physical attacks. The HSM consists of a CPU, some non-volatile memory, a built-in clock, and some I/O interface. In addition, the HSM has a built-in battery in order to power the clock and the tamper detection and reaction circuitry.

The main HSM functions include cryptographic operations, as well as key and device management functions. The main cryptographic operations provided by the HSM are the digital signature generation and the decryption of encrypted messages. The digital signature
generation function is mainly used by the secure communication module (see Sec.~\ref{sec:sec-comm}) for signing outgoing messages. The HSM always includes a timestamp in every signature that it generates, which makes it possible to detect replay attacks. The decryption function is mainly used by the pseudonym handling application, which receives the anonymous certificates in an encrypted form from the pseudonym provider.

The HSM handles short-term keys for the short-term identification and long-term keys for the long-term identification of the vehicle. These keys are generated by the HSM, and only the public keys are output from the device. The generation of short-term keys can be
initiated by any application running on the OBU. In contrast, the long-term keys are generated at manufacturing time, however, they can be updated later by trusted authorities.

Device management and long-term key update are achieved through signed commands from the CA. In order to verify the signature on these commands, the HSM stores trusted root public keys that are loaded into the device during the initialization procedure in a secure environment. We envision two such \emph{root public keys}, $K_{1}$ and $K_{2}$, in the HSM, with the corresponding private keys held by the CA. In case one of the CA'a private keys is compromised, the corresponding public key, say $K_{1}$, can be revoked, as discussed in the next paragraph. The revocation command must be signed with the private key corresponding to $K_{1}$ itself. Once $K_{1}$ is revoked, a new key $K_{1}'$ can be loaded into the HSM by a command signed with the private key corresponding to $K_{2}$. In addition, when $K_{1}$ is revoked, the HSM does not accept commands aimed at revoking $K_{2}$. This scheme ensures secure root key update unless both root keys are compromised.

As discussed next, CA commands can include revocation of the entire device. The revocation of the HSM is achieved by a signed kill command, which deletes every piece of information from the memory, making the device unusable. Further device management functions include device initialization, and clock synchronization. During device initialization, the main parameters of the HSM, as well as the root public keys are loaded in the HSM. Clock synchronization allows for synchronizing the internal clock of the HSM to a trusted external clock.

\begin{figure}[t]
\begin{center}
\includegraphics[width = \columnwidth]{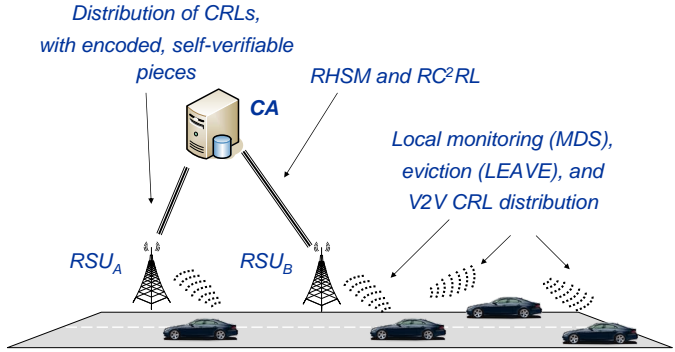}
\caption{Solutions of the Revocation Problem in VC Systems.} \label{fig:revocation}
\end{center}
\end{figure}

\subsection{Revocation}

The certificates of faulty nodes have to be revoked, to prevent them from causing damage to the VC system. Revocation can be decided by the CA because of administrative or technical reasons. The basic mechanism to achieve this are \emph{Certificate Revocation Lists}  (CRLs) the CA creates and authenticates. The challenge is to distribute effectively and efficiently the CRLs, which can be achieved by a combination of methods illustrated in Fig.~\ref{fig:revocation}.

We leverage on the road-side infrastructure to distribute CRLs. We find that with RSUs placed on the average some kilometers apart, and with CRL distribution by each RSU at a few kbps, all vehicles can obtain CRLs of hundreds of kilobytes over a time period of an average commute \cite{PapMH2008}. This is achieved primarily due the use of encoding of CRLs into numerous (cryptographically) self-verifiable pieces and low-rate broadcast transmission of CRL pieces. In areas with no RSUs, V2V CRL distribution initiated by vehicles that were previously in contact with RSUs, or use of other communication technologies, could have a complementary role. The size of CRLs and the overall amount of revocation information to be distributed can still be a challenge. At first, collaboration between CAs, so that CRLs contain only regional revocation information, can keep the CRL size low~\cite{PapMH2008}.

Revocation can leverage on the HSM, with the CA initiating the RHSM (Revocation of the HSM) protocol \cite{revocation}, issuing a ``kill'' command signed with the private key corresponding to one of the root public keys. If a HSM receives a kill command, it deletes everything from its memory including its own private keys, to prevent the generation of any new keys or signatures by the compromised module. The CA determines the location of the vehicle and sends the kill command via the nearest RSU(s). The HSM has to confirm the reception of this command by sending an ACK before erasing the long term signature generation key $(\mathit{SK}_X)$. If communication via the RSUs fails (i.e., an ACK is not received after a timeout), the CA can broadcast the command via the RDS (Radio Data System).

If the adversary controls the CA-HSM communication, the CRL-based revocation has to be performed. This can also be done via the RC$^2$RL (Revocation using Compressed Certificate Revocation Lists) protocol \cite{revocation}, which can reduce the size of CRLs  by a lossy compression scheme, notably Bloom filters, to the extent they could be transmitted even over the RDS. The identification of a revoked certificate in the Bloom filter is always possible (zero false negative rate), along with a configurable low false positive rate. An occasional revocation of ``innocent'' credentials, traded-off for compression (efficiency), is not an issue when RC$^2$RL revokes large numbers of short-term credentials.

The inclusion of credentials in a CRL implies that the CA has established the need to revoke the node. If this is because of faulty behavior, the absence of an omni-present monitoring facility makes the detection harder. Moreover, CRLs will be issued rather infrequently (e.g., once per day), thus leaving a vulnerability window until a faulty node is revoked. To address this, we propose that misbehavior detection is left to vehicles, which can then defend themselves by locally voting off and excluding misbehaving vehicles. We propose the use of two localized defense schemes, MDS (Misbehavior Detection System) and LEAVE (Local Eviction of Attackers by Voting Evaluators) \cite{revocation}. The first allows the neighbors of a misbehaving node detect it, and the second enables them to exclude it from the local VC operation. After a LEAVE execution, the evaluators report the misbehaving node to the CA; a node can be revoked by the CA, using one of the previously described approaches, after having been evicted a threshold number of times by its (changing) neighbors.


\section{Secure Communication} \label{sec:sec-comm}


\subsection{Secure Beaconing}


\begin{figure}[t]
\begin{center}
\includegraphics[width = \columnwidth]{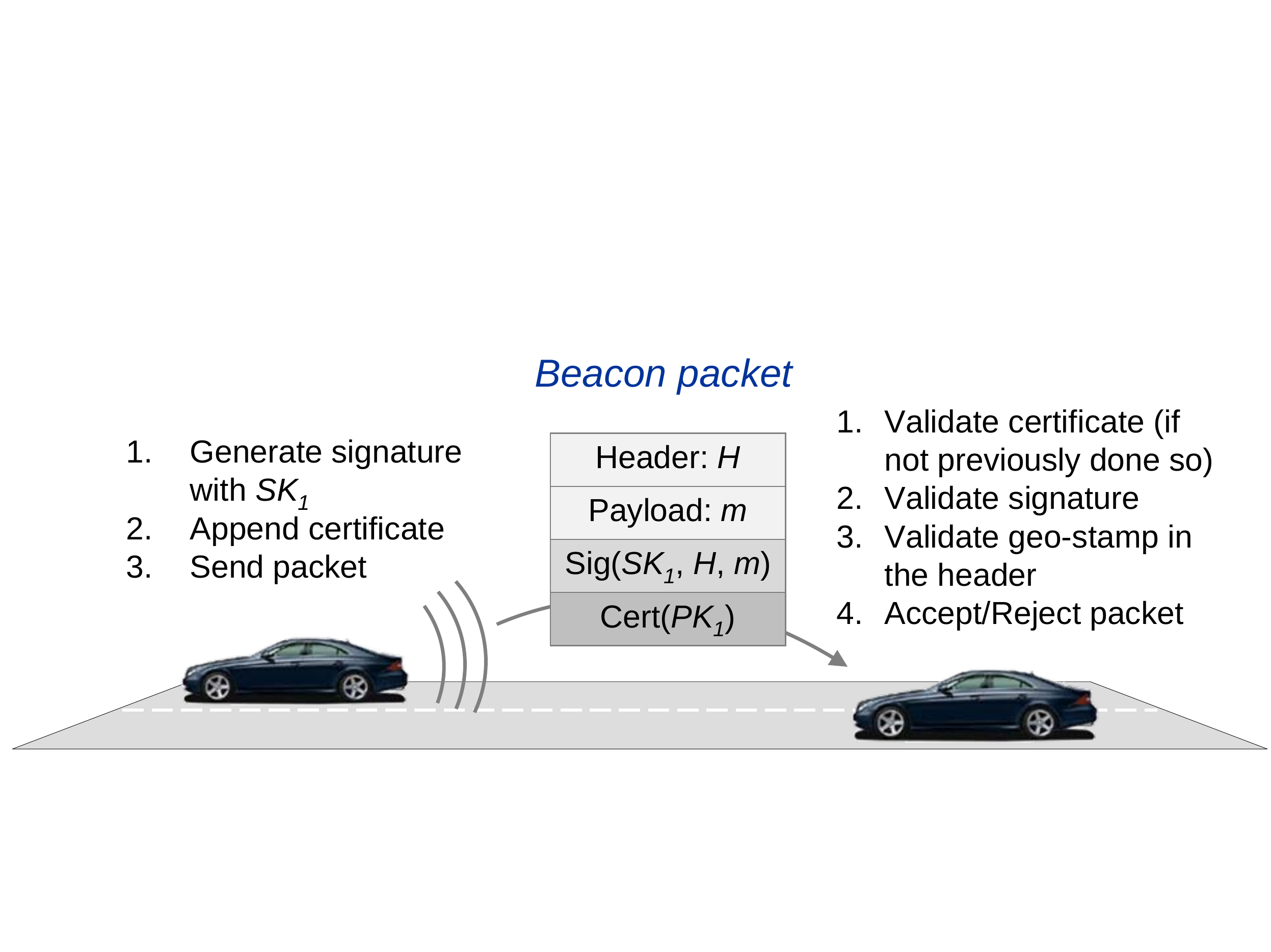}
\caption{Example of Secure Communication: Secure Beaconing}
\label{fig:securebeaconing}
\end{center}
\end{figure}

Beaconing denotes periodic single-hop broadcasts typically used for the so-called Cooperative Awareness applications. In order to create awareness of other vehicles in the vicinity, every beacon contains information on the sender's status such as vehicle position, speed, and heading. The frequency of beacon packets is expected to range from about 10Hz to 1Hz for most use cases.

As already introduced, beacon messages are digitally signed and the signer's certificate is attached. More precisely, after the beacon message assembly is complete and before submitting a message $m$ to the MAC layer for transmission, the sending node ($V$) calculates a signature $sig(m)$, using the private key $SK(j,V)$ corresponding to the $j$-th pseudonym $PK(j,V)$ that is currently in use. $V$ includes a time-stamp and a geographic position at the instant of transmission, together termed as a geo-stamp. Beyond the signature $sig(m)$ that covers all these fields, $V$ also attaches $Cert_A\{PK(j,V)\}$, which attests the validity of $SK(j,V)$. When a beacon message arrives at the receiver, he can verify the message signature using $PK(j,V)$ in the certificate. The attached certificate $Cert_X\{PK(j,V)\}$ can be verified using the pre-installed public key of $CA_A$ (See Fig.~\ref{fig:securebeaconing}).

These measures achieve four goals. First, the receiver of a beacon message can verify its sender is a valid participant of the VC system (either vehicle or RSU). Second, no node can impersonate another node without compromising its HSM. Third, the integrity of the message is protected, as manipulations are detected if the signature is invalid. Finally, the use of the geo-stamp, along with the signature, allows the detection of replay attacks. Details on replay protection mechanisms follow in the next section.





\subsection{Secure Neighbor Discovery}\label{sec:snd}

Cooperative awareness or safety messaging allow vehicles to discover a frequently updated view of other vehicles in proximity, that is, \emph{physical neighbors}. In addition, for the purpose of communication, it is important that vehicles also discover other nodes (vehicles or RSUs) that are directly reachable, that is, their \emph{communication neighbors}. Typically, it is assumed that if two nodes are communication neighbors then they are physical neighbors, and vice versa. However, this is not the case because adversaries mount \emph{relay attacks}, that is, receive and quickly retransmit (replay) messages of remote nodes~\cite{commag08}.



The inclusion of sender time-stamp and location, along with authentication, enable our system to perform provably secure neighbor discovery against \emph{external} adversaries \cite{asiaccs2008}. The basic idea is to estimate the sender-receiver distance based on own coordinates and the location in the received message and time-of-flight (difference between own time and received timestamp). For a protocol-selectable acceptable neighbor range, the receiving node accepts the sender as communication neighbor when the two distance estimates are equal and the sender is authenticated. As a result, vehicles can be ensured that their \emph{neighbor table} includes only nodes that are indeed communication neighbors.

\subsection{Secure GeoCast}



In vehicular communication, the range covered by one-hop beaconing is often not sufficient. Information on certain events such as accidents need to be disseminated in larger areas. This is achieved by Geocast, which comprises three elements: (i) addressing of a geographically defined destination region, (ii) forwarding towards this region and (iii) distribution of the packet within the destination region. Position-based routing, that is, multi-hop, single-path forwarding of packets towards a geographically defined destination, has been shown well-suited to the dynamics of vehicular networks. Position-based routing is realized by greedy routing protocols such as GPSR or CGGC. The distribution of messages among all nodes within the destination region can be done by simple flooding or by more efficient approaches to multi-hop broadcast. In case of simple flooding, every node inside the destination region rebroadcasts the message once and records its sequence number to suppress re-broadcasting of the same message.

As a basic security measure for both position-based routing and message distribution, source nodes sign created messages and attach the corresponding certificate, similarly to the functionality for Secure Beaconing. Moreover, forwarding nodes can also sign packets they relay, so that they can be authenticated by the next-hop relay \cite{Harsch:07:SecureGeoRouting}. This way, only qualified network participants can create messages that are accepted by others and message integrity is protected towards the destination. Replay and neighbor discovery attacks can be prevented, as discussed in the previous section.

Since beaconing is the basis for position-based forwarding decisions, the location given in beacons can be forged, with data delivery failures (when traffic is attracted by the attacker), and increased network load (due to routing loops). We propose a position verification approach based on plausibility heuristics that is capable to detect such position falsifications~\cite{LeinmuellerSchochKargl:06:LocVerificationVANET}. Second, changing pseudonyms for privacy reasons leads to increased instability in nodes' neighbor tables. This can result in transmission faults to the next hop, because a node is not reachable any more after a pseudonym change, which deteriorate routing performance~\cite{Schoch:06:PseudonymChanges}. To balance network and privacy needs, we can extend routing mechanisms by a MAC layer callback that notifies the routing layer about missed neighbors. Finally, to mitigate resource depletion attacks, with an internal adversary distributing at high rates messages across a large destination region, we propose rate limiting.





\subsection{Pseudonym Handling}\label{sec:pnyms}


An adversary analyzing which certificate are attached to signed messages can track the location of vehicles over time. Hence, we propose to load each vehicle with multiple certified public keys (i.e., pseudonyms) that it uses for short periods of time. If pseudonyms are changed at appropriate time and location, messages signed under different pseudonyms are hard to be linked by an adversary.

As the adversary could use information from other layers of the communication stack to track vehicles (e.g., MAC, IP), a change of pseudonym should be accompanied by a change of the vehicle identifiers in underlying protocols as well. Still, using the location contained in messages to match pseudonyms, an adversary can indirectly identify vehicles by predicting the next position of vehicle even if a vehicle has a new pseudonym. Cloaking of location information~\cite{Gruteser2003a} is not a solution as it would jeopardize the use of safety applications. In~\cite{Buttyan2007a}, we propose that vehicles change pseudonyms in regions not monitored by an adversary. These regions are called mix zones~\cite{Beresford2003a} as the vehicles by changing pseudonyms will mix with each other. We also suggest that vehicles change their pseudonym at regular intervals maximizing the probability of changing pseudonym in a mix zone. In~\cite{Freudiger2007a}, we explore another approach that creates un-monitored regions by encrypting communications (i.e., cryptographic mix zones) in small regions with the help of the road infrastructure.

\begin{figure}[t]
\begin{center}
\includegraphics[width = \columnwidth]{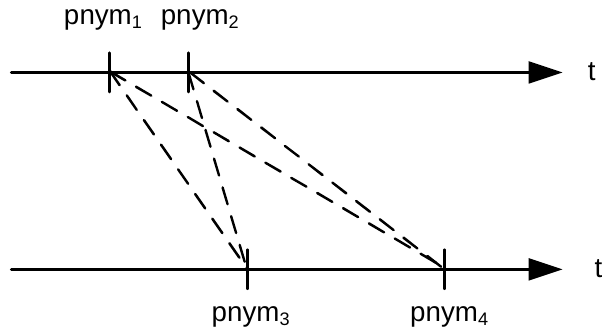}
\caption{Privacy Protection: Given vehicles entering and exiting a mix zone with different pseudonyms, the adversary must predict the most likely matching of event.}
\label{fig:mixzone}
\end{center}
\end{figure}

The general idea of mix zones is explained as follows: If only one vehicle changes its pseudonym in a mix zone, an adversary observing vehicles entering and exiting the region can trivially track vehicles as only one pseudonym changed. But if more than one vehicle changes pseudonym in a mix zone, the adversary must consider every possible matchings of entering and exiting vehicles and estimate the most likely matching given its belief about the mobility of vehicles, the time to traverse the mix zone and the geometry of the mix zone (Fig.~\ref{fig:mixzone}). An adversary will thus find several possible matchings weighted with different probabilities. To measure the amount of location privacy achieved by vehicles in the mix zone, we capture the uncertainty of the adversary with the notion of entropy as defined in~\cite{Serjantov2002a}. With this metric, if all possible matchings look equally likely to an adversary, (i.e. the mix zone is very unpredictable), the adversary is highly inefficient in tracking vehicles. In general, the more vehicles are in a mix zone, the more difficult it is for an adversary to obtain a good estimation of probabilities as many combinations are possible. If the mobility of the vehicles is so that they are equally likely to enter/exit the mix zone from any road, it is difficult for the adversary to obtain precise predictions.

When vehicles change pseudonyms in un-monitored regions of the network, then mix zones are large and it is difficult for an adversary to obtain good estimations. However, when mix zones are created by the use of cryptography, then they tend to be smaller, and thus must be located appropriately to maximize their effectiveness (e.g., at traffic intersections). Hence, linking messages signed under different pseudonyms becomes increasingly hard over time and space for an adversary. As vehicles will change pseudonyms several times before reaching destination, the adversary will accumulate more uncertainty and like in mix network \cite{Serjantov2002a}, mobile nodes can achieve a high level of location privacy.

\section{Conclusions}

We have developed a security architecture for VC systems, aiming at a solution that is both comprehensive and practical. We have studied the problem at hand systematically, identifying threats and models of adversarial behavior as well as security and privacy requirements that are relevant to the VC context. We introduced a range of mechanisms, to handle identity and credential management, and to secure communication while enhancing privacy. In the second paper of this contribution, we discuss implementation and performance aspects, present a gamut of research investigations and results towards further strengthening secure VC systems and addressing remaining research challenges towards further development and deployment of our architecture.  

\bibliographystyle{plain}
\bibliography{refs}

\end{document}